# Millimeter-wave Evolution for 5G Cellular Networks

Kei SAKAGUCHI†a), Gia Khanh TRAN††, Hidekazu SHIMODAIRA††, Shinobu NANBA†††, Toshiaki SAKURAI††††, Koji TAKINAMI†††††, Isabelle SIAUD*, Emilio Calvanese STRINATI**, Antonio CAPONE***, Ingolf KARLS****, Reza AREFI****, *and* Thomas HAUSTEIN*****

**SUMMARY** Triggered by the explosion of mobile traffic, 5G (5th Generation) cellular network requires evolution to increase the system rate 1000 times higher than the current systems in 10 years. Motivated by this common problem, there are several studies to integrate mm-wave access into current cellular networks as multi-band heterogeneous networks to exploit the ultra-wideband aspect of the mm-wave band. The authors of this paper have proposed comprehensive architecture of cellular networks with mm-wave access, where mm-wave small cell basestations and a conventional macro basestation are connected to Centralized-RAN (C-RAN) to effectively operate the system by enabling power efficient seamless handover as well as centralized resource control including dynamic cell structuring to match the limited coverage of mm-wave access with high traffic user locations via user-plane/control-plane splitting. In this paper, to prove the effectiveness of the proposed 5G cellular networks with mm-wave access, system level simulation is conducted by introducing an expected future traffic model, a measurement based mm-wave propagation model, and a centralized cell association algorithm by exploiting the C-RAN architecture. The numerical results show the effectiveness of the proposed network to realize 1000 times higher system rate than the current network in 10 years which is not achieved by the small cells using commonly considered 3.5 GHz band. Furthermore, the paper also gives latest status of mm-wave devices and regulations to show the feasibility of using mm-wave in the 5G systems.
***key words:*** *Millimeter-wave, 5G cellular network, small cell, C-RAN, system rate gain*

a) E-mail: sakaguchi@comm.eng.osaka-u.ac.jp

## 1. Introduction

Due to the popularization of smart phones and tablets in recent years, the traffic load on conventional cellular networks is predicted to be increased by 1000 times in the next 10 years [1]. To face the severe issue of system capacity shortage due to the increasing data traffic in cellular networks, standardization on heterogeneous networks (HetNet) with overlay deployment of low-power basestations (BSs) in the service area of conventional ones is being done by the 3GPP (3rd Generation Partnership Project) – an international standardization body of cellular networks [2]. Another significant contributor to increase capacity is spectrum extension, e.g. 3.5 GHz band, which was prepared by ITU-R (International Telecommunication Union Radio communication sector) in WRC-07 (World Radiocommunication Conference) [3] as a common band in the world, provides 100 MHz bandwidth for both downlink and uplink to achieve 10 times higher data rate than the current 3GPP LTE (Long Term Evolution) [4].

With the objective of bandwidth expansion, mm-wave band is much attractive since ultrawide bandwidth is available as shown in Sect. 5.3 in this paper, e.g. up to 7 GHz of continuous spectrum is available worldwide at the 60 GHz unlicensed band. It is also noted that the mm-wave band provides preferable propagation characteristics of high path loss and high oxygen absorption which helps in reducing interference between neighboring connections. Besides, applying more bandwidth per communication link is a significant contributor for improved energy efficiency measured in Joules-per-transmitted-bit which contributes to the birth of green ICT (Information Communications Technology) networks. Additionally, monolithic integrated circuits are already available on a large scale basis with the advent of the 60 GHz extension of Wi-Fi in IEEE 802.11ad standard [5, 6]. Despite such advantages of mm-wave, current standardized mm-wave communication systems such as WiGig and 802.11ad, are not integrated into cellular networks yet. Their applications are still restricted to only performing data transfer between AV (Audio Visual) equipment or working as a bridge for Internet connection across buildings.

Regarding the evolution of current LTE or LTE-Advanced to its future generation to solve the problem of capacity shortage, some researchers and industrial experts have started to investigate the potential of mm-wave cellular networks [7, 8, 9, 10, 11]. R. Heath and his research team provided initial theoretical results on the capacity and coverage of cellular networks using mm-wave [8]. T. Rappaport and his research team have been pushing mm-wave bands for 5G cellular networks with evidences of mm-wave propagation measurements [9]. A. Ghosh and his industrial team proposed to use mm-wave beamforming both for access and backhaul in smallcell networks [10]. W. Roh and his industrial team revealed the effectiveness of mm-wave beamforming to improve system capacity of future cellular networks [11]. All of these works are very important to support mm-wave in 5G cellular networks, however, the analyses in these papers are limited to link level or system level with homogeneous networks. Beside theoretical works, T. Rappaport led his research team to conduct mm-wave propagation measurements at 28 GHz and 38 GHz [9] in both indoor and outdoor environments. The experimental results showed significant potential of introducing mm-wave into cellular networks with the coverage up to 200 m in some specific scenarios. His team also extended the frequency up to 73 GHz in [12]. Furthermore, [13] developed a prototype hardware of mm-wave beamforming with 32-element uniform planer array to show the potential of mm-wave for 5G cellular networks. Despite these theoretical and experimental works showed high potential of mm-wave link, they lacked a systematic point of view of how to integrate mm-wave into current cellular networks efficiently. Comprehensive system level analysis is also needed to fully understand the effectiveness of mm-wave integrated HetNets.

This paper presents our pioneering works [14] in evolving mm-wave into future 5G cellular networks and aims to extend the network capacity by 1000 times. In [15, 16], we have proposed a novel architecture of mm-wave overlay HetNet in which mm-wave ultra-wideband BSs employing recent state-of-the-art technologies of mm-wave devices are introduced and integrated into conventional cellular networks such as LTE. As in other similar proposals [17, 18, 19] for 5G systems, we introduced the concept of C-RAN (Centralized RAN) [20, 21, 22] and U/C (User-plane/Control-plane) splitting [23] in our architecture. Since the coverage of mm-wave access is limited, cloud cooperation by the C-RAN using C-plane via legacy cellular networks is indispensable to effectively operate HetNet with mm-wave access. More precisely, this architecture enables centralized resource control, e.g. centralized cell association, efficient cell discovery and cooperative beamforming, that boosts the performance of mm-wave access employing ultra-wideband high gain directional antennas. Different from the previous works [15, 16, 17, 18, 19], this paper gives systematic performance results of the mm-wave overlay HetNet operated with C-RAN and U/C splitting. To show the realistic performance, this paper introduces a future traffic model estimated from current traffic data, a mm-wave propagation model derived from measurement campaign, and a novel centralized cell association algorithm for multi-band HetNets. Numerical simulation reveals interesting results that the proposed system with mm-wave access realizes 1000 times system rate in the next 10 years, which cannot be realized by other schemes such as using 3.5 GHz access. Moreover, to show the feasibility of the proposed system, this paper also presents the latest state-of-the-art mm-wave devices.

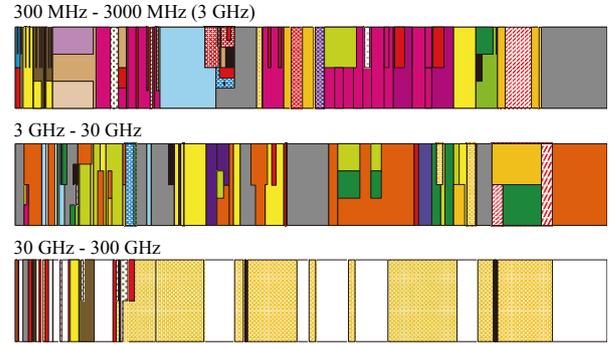

**Fig. 1**  Frequency spectrum allocation of Japan in year 2014.

The paper is organized as follows. Section 2 explains the necessity of introducing high frequency and small cell in future 5G cellular networks. Core technologies in millimeter-wave access for 5G cellular networks are reviewed in Sect. 3. Prospective of system performance is evaluated by introducing an expected future traffic model, a measurement based mm-wave propagation model, and a centralized cell association algorithm exploiting the C-RAN architecture is presented in Sect. 4. To show the feasibility of the proposed system, Sect. 5 presents current status of millimeter-wave devices and regulations. Finally, Sect. 6 concludes the paper and suggests future directions.

## 2. Why High Frequency and Small Cell

Let us go back to Shannon's capacity theorem to explain why high frequency and small cell are effective to increase data rate.

2.1 Frequency and Data Rate

Assuming a data transmission system with its bandwidth of $B$ and received SNR (signal-to-noise ratio) of $\gamma$, the Shannon's capacity theorem gives achievable upper bound of data rate $C$ in bps with error free as

$$C = B \log_2(1+\gamma). \qquad (1)$$

In this equation, the received SNR $\gamma$ is the ratio between the received signal's power $P_r$ and the receiver noise $P_n$, and is given as follows

$$\gamma = \frac{P_r}{P_n} = \frac{P_r}{BN_0}, \qquad (2)$$

where $N_0$ is the power spectral density of the noise.

In the case of radio communication systems, the bandwidth is a function of its center frequency $f_0$ i.e. $B = \alpha f_0$ due to radio regulations and limitation of RF (radio frequency) circuits, where $\alpha$ is called fractional bandwidth and its typical value is $\alpha = 1\%$ in recent radio systems. In wireless communication systems, coverage of the system is defined or controlled by a minimum required SNR $\gamma_0$. To keep $\gamma_0$ constant, the corresponding received power $P_{r0}$ should be a function of $f_0$ as follows,

$$P_{r0} = \gamma_0 \alpha f_0 N_0. \qquad (3)$$

If the condition on $\gamma_0$ is satisfied at the coverage edge, Eq. (1) indicates that the achievable data rate $C$ is a linear function of the center frequency $f_0$. Therefore, if $B = \alpha f_0$ is satisfied, the higher the frequency is the higher data rate can be realized. Figure 1 shows current frequency spectrum allocation in Japan [24]. Since it is drawn in logarithmic scale, it is obvious that the available bandwidth is wider in higher frequency. This paper prefers to utilize these higher frequencies to boost the system capacity of cellular networks.

2.2 Frequency and Coverage

Next, let us discuss about coverage of radio systems in terms of frequency. Based on the basic Friis equation, the received power of the system in a free space can be modeled by using $f_0$ as

$$P_r = \left(\frac{c}{4\pi d f_0}\right)^2 G_r G_t P_t, \qquad (4)$$

where $P_t$ is the transmit power, $G_t$ and $G_r$ are gains of the transmit and receive antennas respectively, $d$ is the distance between the transmitter and the receiver, and $c$ is the speed of light. The coverage $d_0$ can be defined as the distance which realizes the minimum SNR $\gamma_0$. By substituting Eq. (3) into Eq. (4), the coverage of the system is calculated as follows

$$d_0^2 = \frac{c^2}{\gamma_0 \alpha N_0 (4\pi)^2} G_r G_t P_t \frac{1}{f_0^3} = \chi \frac{1}{f_0^3} \qquad (5)$$

where $\chi$ is a constant related to the minimum SNR.

This equation implies that in free space or in a propagation environment with path loss exponent $\beta = 2$ like mm-wave band, the coverage becomes

$$d_0 = \chi f_0^{-\frac{3}{2}}, \qquad (6)$$

and in a general case with the parameter $\beta$ it becomes

$$d_0 = \chi f_0^{-\frac{3}{\beta}}. \qquad (7)$$

For example with $\beta = 3$ as in typical urban environment, the coverage is inversely proportional to the frequency. So that there is a tradeoff between data rate and coverage with respect to the frequency. It means, if $B = \alpha f_0$ and $\beta = 3$ are satisfied, the data rate is proportional to the frequency while the coverage is inversely proportional.

In the case of mm-wave, for a fixed antenna aperture, higher antenna gain can be achieved at higher frequency. For example, if a linear array antenna is employed at the transmitter, the gain of the transmit antenna becomes a function of $f_0$ as well since the number of antenna elements is linearly increased by decreasing the wavelength. Therefore, the gain of antenna $G_t$ can be written as

$$G_t = \delta f_0, \qquad (8)$$

where $\delta$ is a constant value. By substituting Eq. (8) into Eq. (5), the coverage becomes inversely proportional to the frequency in this case as well. Therefore, the equation of $d_0 = \chi/f_0$ is general for both current microwave band with $\beta = 3$ and future mm-wave band employing beamforming antennas with $\beta = 2$ and $G_t = \delta f_0$. It is noted that if a planar-typed array antenna is introduced at the transmitter, the gain $G_t$ becomes a square function of frequency, thus the coverage reduction can be further relaxed as $d_0 = \chi f_0^{-1/2}$.

2.3 Frequency and User Rate

In the final step, let us calculate the data rate per user by assuming a single BS is accessed by multiple users. By assuming the number of users located in the coverage as $N_{UE}$, the user data rate $C_{UE}$ in bps/user is calculated as follows assuming

Table 1  Examples of multi-band HetNet.

| Macro BS | Center frequency | 2 GHz |
|---|---|---|
| | Bandwidth | 10 MHz |
| | Tx power | 46 dBm |
| Smallcell BS (3.5 GHz) | Center frequency | 3.5 GHz |
| | Bandwidth | 100 MHz |
| | Tx power | 30 dBm |
| Smallcell BS (60 GHz) | Center frequency | 60 GHz |
| | Bandwidth | 2.16 GHz |
| | Tx power | 10 dBm |

- orthogonal multiple access.

$$C_{UE} = \frac{B \log_2(1+\gamma)}{N_{UE}} \qquad (9)$$

The number of users $N_{UE}$ can be calculated using $d_0$ as follows,

$$N_{UE} = \pi d_0^2 \eta, \qquad (10)$$

where $\eta$ in user/m² is the user density. Finally, substituting the assumptions $B = \alpha f_0$ and $d_0 = \chi/f_0$ into Eq. (9) and Eq. (10), we achieve

$$C_{UE} = \frac{\alpha \log_2(1+\gamma)}{\pi \eta \chi^2} f_0^3 = O(f_0^3) \qquad (11)$$

This equation leads us to the era of high frequency and small cell. For example, this equation indicates that 10 times frequency achieves 1000 times system rate. It is also noted that, the density of small coverage BSs should be increased with the order of $O(f_0^2)$ to achieve system rate gain with the same order of Eq. (11).

Summarizing these discussions, the following proposition can be stated. In 5G cellular systems, 1000 times system rate is achieved by at least 10 times bandwidth and 100 times small coverage BSs.

## 3. Millimeter-wave Access for 5G Cellular Networks

Motivated from the discussions in the previous section, the authors have been studying a new HetNet architecture with high frequency and small cells for future 5G cellular networks [15]. We will review the proposed architecture in this section.

### 3.1 Multi-band HetNet

HetNet, which is a new type of cellular network topology constructed by mixture of the conventional large coverage macro BS and low power with small coverage BSs (smallcell BSs) was proposed to offload user traffic to small cells. Not only the user rates at the vicinity of smallcell BSs are increased, but also the overall system rate can be improved by decreasing the traffic load on the macro BS by offloading.

In the conventional single-band HetNet standardized in 3GPP Release 10 [2], the same band is used both for the macro BS and smallcell BSs. Therefore, the single-band HetNet requires macro-smallcell interference mitigation techniques such as ABS (almost blank subframe) [2] in 3GPP standard. However, since the available bandwidth is split in time or frequency domain in interference control schemes, channelization loss occurs in the conventional HetNet.

In recent standard of 3GPP Release 12 [25], multi-band HetNet with inter-site carrier aggregation capability has been standardized where macro and smallcell BSs use different frequency bands. Therefore, macro-smallcell interference control schemes are not needed anymore, however, development of new user equipment (UE) with dual connectivity for two different bands is needed. 3.5 GHz band, which is legislated by ITU-R, has high potential to be used for smallcell BSs with an available bandwidth of 100 MHz. Moreover in the future 5G systems, much higher frequency such as 60 GHz band should be included for access link. Table 1 lists up examples of multi-band HetNet. In this example, 2 GHz band is used for macro BS, while 3.5 GHz or/and 60 GHz bands are considered for smallcell BSs. 3.5 GHz band can offer a 10 times bandwidth compared to the current system, while 60 GHz band offers more than 100 times bandwidth, which is definitely attractive for the future 5G systems.

However, there are several drawbacks in the multi-band HetNet, since the small cells are operated at different frequency band from macro BS and their coverage is not continuous and scattered in the macro cell. From UE point of view, the UEs are required to support dual connectivity to realize multi-band HetNet. The first problem is power consumption of UE to find smallcell BS by running cell search all the time when the UE is connected to the macro BS. The second problem, from network perspective, is handover failure. Since coverage of small cell is limited, it is not effective to perform regular handover process as in the conventional scenario of e.g. two macro BSs. The third problem, from BS point of view especially in the case of 60 GHz smallcell BSs, is the mismatch between the coverage of small cells and location of UEs. Since the number of UEs in the coverage of small cell is relatively small compared to the conventional macro cell, dynamic cell structuring technology is necessary to operate multi-band HetNet effectively.

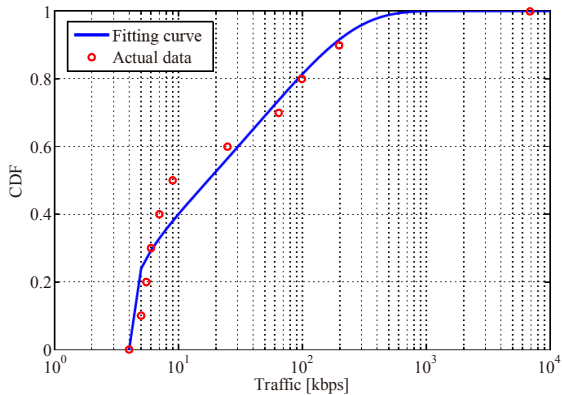

Fig. 3  CDF of measured and modeled user traffic distributions.

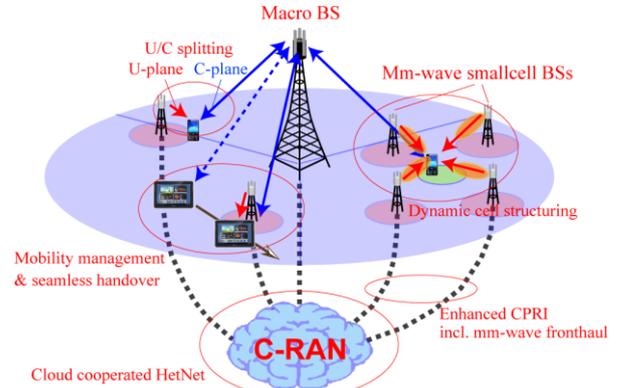

Fig. 2  Proposed architecture for 5G cellular networks.

Table 2  Gamma distribution parameters.

| | | |
|---|---|---|
| Shape parameter | $k$ | 0.2892 |
| Scale parameter | $\theta$ | $2.012 \times 10^5$ |
| Traffic bias | | 4 kbps |

3.2 Proposed Architecture for 5G Cellular Networks

Figure 2 shows our proposed architecture for 5G cellular networks. In this architecture, not only the 3.5 GHz smallcell BSs but also 60 GHz mm-wave smallcell BSs are introduced and overlaid in the conventional macro cell to form a multi-band HetNet. To overcome the issues presented in Sect. 3.1, the concept of cloud cooperation is introduced. In the cloud cooperated HetNet, all smallcell BSs as well as macro BS are connected to C-RAN. If all macro and smallcell BSs are employing 3GPP LTE standard, it is easy to develop such kind of architecture by introducing RRHs (remote radio heads) for smallcell BSs and connecting them with C-RAN with CPRI (common public radio interface) [26]. On the other hand, if we employ WiGig standard for mm-wave smallcell BSs, there is no off-the-shelf interface between C-RAN and smallcell BSs. However, we believe that such kind of interface can be developed by extending existing protocols such as CAPWAP (the control and provisioning of wireless access point) [27] in the framework of 3GPP/Wi-Fi interworking. In this paper, we call this future interface including legacy CPRI and recent ETSI ORI (open radio equipment interface) standard [28] as enhanced CPRI. This kind of architecture is suitable for introducing even a large number of smallcell BSs into the macro cell as will be shown in Sect. 4, since C-RAN controls all smallcell BSs and UEs based on the measurement and report given from the macro BS. It is noted that the network topology is not necessarily star type but cluster tree type is also possible where the number of fronthaul links connected to C-RAN is reduced. Moreover, wireless fronthaul between C-RAN and smallcell BSs is almost ready at mm-wave band [29] that will reduce the cost of deployment of a large number of smallcell BSs.

Based on the C-RAN architecture, a concept of U/C splitting is introduced where macro BS manages C-plane of all users while U-plane can be connected to smallcell BSs opportunistically. This is heterogeneity on role where macro BS works on C-plane to guarantee connectivity while smallcell BS works on U-plane to provide high data rate. This strategy enables the macro BS to manage mobility and traffic of all users in the HetNet in a centralized manner. So that the macro BS can assist UEs for cell discovery, which is partially supported in current ANDSF (access network discovery and selection function) [30], and also assist seamless handover via C-RAN. As in the other proposals for 5G systems, the technology of U/C splitting is inevitable for operating multi-band HetNet. Especially in the case with mm-wave smallcell BSs, this technology is mandatory since the coverage of mm-wave cell is extremely limited.

The C-RAN architecture and mobility/traffic management using C-plane via macro BS enable centralized resource control such as cell association to maximize system rate as explained in Sect. 4. Moreover, since the mobility information is available in C-RAN, dynamic cell structuring or virtual cell can be realized which is important to overcome the problem of limited coverage especially in the case of mm-wave smallcell BSs. In the dynamic cell structuring, cell structure of small cells are dynamically controlled to track high traffic users or hotspots by means of beamforming antenna and power control via C-RAN. Generally, beamforming technology is effective to compensate for distance dependent path loss while it also creates hidden terminals located outside of the main beam even at the vicinity of smallcell BS. If the C-plane of these hidden terminals is disconnected, it is impossible to steer the beam of smallcell BS to them. On the other hand, since the C-plane is managed by macro BS in this architecture, the hidden terminal problem does not occur and dynamic cell structuring using beamforming technology can work effectively. Such dynamic cell structuring including the concept of smallcell BS dormancy (switch on/off) will also contribute to save power consumption if the high traffic user moves away from the vicinity of smallcell BSs. Moreover, cooperative transmission among multiple smallcell BSs can also be performed since all the smallcell BSs are connected with C-RAN. For example, cooperative scheduling/beamforming is very effective to combat shadowing problem typically occurred in mm-wave band. Furthermore, the cooperative transmission by a large number of smallcell BSs with beamforming is very effective to support hotspot users located in-between the smallcell BSs [31]. This kind of dynamic cell structuring is considered to transfer the wasted radio resources in sparse areas to congestion area via cooperative beamforming of smallcell BSs.

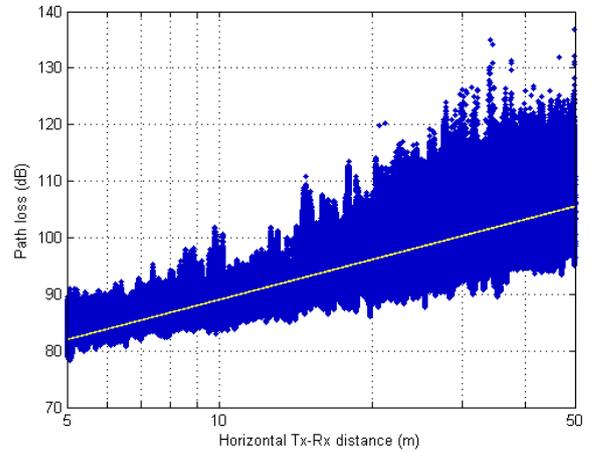

**Fig. 4** Measured path loss and mean least squares fit.

## 4. Performance Perspective

In this section, the performance of the proposed 5G cellular network with mm-wave access is evaluated via system level simulation. To evaluate the realistic performance, a new traffic model, a measurement based mm-wave propagation model, and a novel centralized cell association algorithm are introduced in this section. The new traffic model is created by using actual traffic data measured in a dense urban area in Japan, and future traffic is predicted based on the recent growth rate of mobile traffic. Since there is no standard mm-wave channel model for outdoor access scenarios, a new mm-wave channel model based on an outdoor measurement campaign [32] is applied in this paper. Moreover, as a first step of the centralized resource management by C-RAN, a new cell association scheme [33] proposed by the authors is also introduced for multi-band HetNet. Finally, the system rate gain are calculated and compared for two types of multi-band HetNet with 3.5 GHz and 60 GHz small cells.

4.1 Traffic Model

In order to evaluate the system rate gain accurately, the model of user traffic demand reflecting actual environment should be introduced. We employ Gamma distribution with certain bias to describe the instantaneous user traffic demand of the actual traffic data in a dense urban area. This user traffic data were measured during the daytime in the most populous part within 23 wards, Tokyo, Japan in 2013. It contains all kinds of mobile traffic data sent through currently operated macro BSs. The Gamma distribution is defined as:

$$f(x) = x^{k-1} \frac{\exp(-x/\theta)}{\Gamma(k)\theta^k} \qquad (12)$$

where $k$ is a shape parameter, $\theta$ is a scale parameter, and $\Gamma(\cdot)$ is a Gamma function. Based on the actual traffic data, the average packet generation interval per user was 8 seconds and the CDF of packet size had long-tail characteristic. By dividing the packet size with average packet generation interval and fitting it with Gamma distribution, we obtain the distribution of the instantaneous user traffic demand. Figure 3 shows the CDF of the measured and fitted distributions and Table 2 shows the derived shape parameter, scale parameter, and the traffic bias describing the minimum traffic load of the environment. By assuming that shape parameter will not change in the future, average traffic value can be changed by controlling the scale parameter. In this paper, assuming that the mobile traffic load grows about twice every year, two scenarios are evaluated at present and 10 years later (1000 times higher traffic).

4.2 Measurement based mm-wave channel model

In order to evaluate the performance of mm-wave small cell access, a measurement based mm-wave channel model [32] is introduced in this paper. A measurement campaign in an outdoor small cell access environment has been performed in the Potsdamer Straße in downtown Berlin, Germany which was selected as a typical densely built and busy environment. Such places with a large number of pedestrians are expected to be among the first places that will be equipped with mm-wave small cell base stations.

The transmitter was mounted on a pole at a height of 3.5 meter to simulate a typical small cell deployment on street furniture, such as street lights. The receiver was mounted on a mobile cart at a height of 1.5 meter and moved along the sidewalk at constant speed. This should reflect a typical mobile user terminal.

Based on the results of the measurement campaign a specific path loss model was derived. This model does not separate large-scale and small-scale behavior explicitly, where the parameters of small-scale path loss depend also on the distance.

In Fig. 4 the path loss for all distances from 5 to 50 meter distance is shown as a scatter plot (blue). The yellow line represents the linear least squares (LS) fit corresponding to the mean path loss drawn by the following equation where $d_{ref}$ corresponds to the reference distance of 5m.

$$PL(d) = \begin{cases} 82.02 + 23.6 \log_{10}\left(\dfrac{d}{d_{ref}}\right), & d \geq d_{ref} \\ 82.02 + 20 \log_{10}\left(\dfrac{d}{d_{ref}}\right), & d < d_{ref} \end{cases} \quad (13)$$

To further investigate the path loss in a statistical manner, the data was normalized with the fitted mean path loss. Then the relative amplitude of each snapshot was calculated and empirical probability distributions for bins of 5 meter distance were generated. For each of these distributions, a Rician distribution was fit onto the data, which has been identified to be well suited for this purpose. To simplify the model, a linear fit has been performed on the fitted parameters of the Rician factor $K(d)$ and mean power $\Omega(d)$ as follows.

$$K(d) = -0.62d + 25 \, \text{dB} \quad (14)$$

$$\Omega(d) = 0.013d + 1 \quad (15)$$

Based on these parameters, channel coefficients corresponding to the small-scale path loss (fading) are generated based on a non-central complex normal distribution with the variance of $\Omega(d)$ and mean of $\dfrac{K(d)}{1+K(d)}$.

4.3 Radio Resource Control

In this study, a new cell association algorithm specialized for multi-band HetNet is introduced. As for BS deployment, general hexagonal structure with three sector macro cells is assumed as shown in Fig. 5, where the macro BS is located at the center of hexagonal structure and smallcell BSs are overlaid on the macro cells randomly and operated in different frequency bands from macro BSs. It is noted that the proposed cell association algorithm is applied only for U-plane, and it is assumed that the C-plane is always associated to one of macro BSs by using conventional cell association algorithms in advance to the U-plane. In order to maximize the system performance, system rate which expresses the total capacity of the system is introduced. The system rate is defined as:

$$R = \sum_{u \in M} \min\left(\frac{W_M C_{u,M}^e}{|M|}, L_u\right) + \sum_{s=1}^{N_s} \sum_{u \in s} \min\left(\frac{W_s C_{u,s}^e}{|s|}, L_u\right) \quad (16)$$

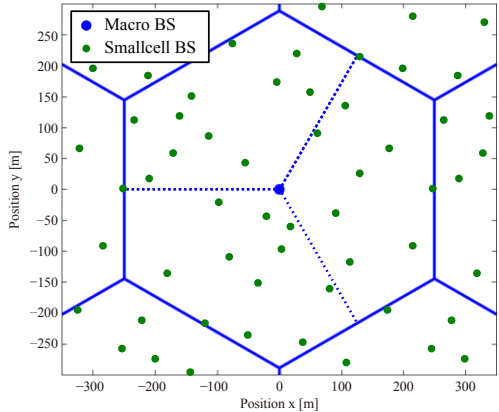

**Fig. 5** BS deployment scenario.

where $W_M$ and $W_s$ are the available bandwidth for macro and small cell respectively. $C_{u,M}^e$ in bps/Hz and $C_{u,s}^e$ in bps/Hz are the received link spectral efficiency of UE $u$ from macro and $s$-th small cell. $|M|$ and $|s|$ represent the number of users belong to macro BS and $s$-th smallcell BS respectively. $N_S$ is a total number of smallcell BSs deployed within one macro cell area. $L_u$ in bps is a traffic demand of user $u$. Equation (16) expresses the balance between achievable rate and traffic demand. If achievable rate is much higher than traffic demand, the user rate is peaked by the traffic demand and vice versa.

On the offloading point of view, BS should accommodate appropriate UEs to improve system rate efficiently. In this multi-band HetNet case, it is appropriate if the smallcell BS accommodates the UE whose traffic demand is relatively high. To achieve this purpose, a novel cell association method using combinatorial optimization is introduced.

In homogeneous network and conventional co-channel HetNet, conventional SINR based association, which compares received SINR of all BSs and chooses the best BS, works effectively because the cell edge is common between all neighboring cells. However in the multi-band HetNet case, cell edge of macro and small cells are different. Therefore simple SINR comparison method cannot work effectively. To maximize the system rate in Eq. (16), achievable rate, traffic demand, and the number of users in each cell should be considered simultaneously. The C-RAN based architecture proposed in Sect. 3 is very effective against this kind of centralized optimization problem. Our algorithm based on the combinatorial optimization can satisfy these requirements and it is realized by the following four procedures.

1) Macro and small cell index allocation
2) Problem partition
3) Beamforming (only for mm-wave case)
4) Combinatorial optimization

*1) Macro and Small Cell Index Allocation:* First, macro and small cell indices are allocated to all UEs. These indices indicate the best macro BS and best smallcell BS which can provide larger link spectral efficiency than any other BSs of the same kind. It is noted that the macro cell index indicates the macro BS where C-plane is connected to. This allocation can partition one macro cell area into many small cell areas overlapping with the macro area. These indices are determined based on the link spectral efficiency as follows.

$$\begin{aligned} i_{u,m} &= \arg\max_{m} C_{u,m}^e \\ i_{u,s} &= \arg\max_{s} C_{u,s}^e \end{aligned} \quad (17)$$

*2) Problem partition:* Based on the macro cell area partitioning, the objective function of Eq. (16) can be partitioned into sub-problems corresponding to all small cell areas. This paper assumes that there is no congestion area (hotspot) in terms of both user distribution and traffic as a worst case analysis for HetNet, and smallcell BSs are assumed to be distributed uniform randomly. If scenarios with hotspots are considered, the performance of HetNet is easily improved by introducing smallcell BSs at the vicinity of hotspots as in [34]. By reflecting the effect of this macro cell partitioning, Eq. (16) can be transformed as

$$R = \sum_{s=1}^{N_S} \sum_{u \in M_s} \min\left(\frac{W_M |M_s|}{|M|} \frac{C_{u,M}^e}{|M_s|}, L_u\right)$$
$$+ \sum_{s=1}^{N_S} \sum_{u \in s} \min\left(\frac{W_s C_{u,s}^e}{|s|}, L_u\right)$$
$$= \sum_{s=1}^{N_S} \left(\sum_{u \in M_s} \min\left(\frac{\alpha_s W_M C_{u,M}^e}{|M_s|}, L_u\right) + \sum_{u \in s} \min\left(\frac{W_s C_{u,s}^e}{|s|}, L_u\right)\right)$$
$$= \sum_{s=1}^{N_S} R_s$$
(18)

where $|M_s|$ is the number of macro users within the area of the $s$-th small cell and $\alpha_s = \frac{|M_s|}{|M|}$ is considered as a resource allocation ratio to the macro users in the $s$-th small cell area. This sub-problem $R_s$ can be maximized by combinatorial optimization.

*3) Beamforming:* In the case of mm-wave bands, beamforming technique is introduced in smallcell BSs to compensate distance dependent path loss. This study assumes that vertical tilt angle can be controlled from 0 deg to 180 deg and horizontal tilt angle from 0 deg to 360 deg in steps of 15 deg for beamforming antennas. The beamforming is performed after the index allocation, where the link spectral efficiency is calculated without beamforming because there is no significant difference of the partitioned area between with and without beamforming in the case of the employed beam pattern. This study assumes that all smallcell BSs know the locations of UE within the partitioned area perfectly, so that the BS is able to tune the beam direction towards the scheduled UE.

*4) Combinatorial Optimization:* The important thing in cell association problem in multi-band HetNet is 'who should be accommodated in the small cell?' to maximize $R_s$. Combinatorial optimization can answer this problem. There are several papers applying combinatorial optimization for cell association. However, these papers only consider homogeneous network [35, 36] or cannot obtain a strict solution for optimal user combination since problem relaxation is applied [37, 38, 39, 40]. On the other hand, our method can be applied to multi-band HetNet and also provides strict user combinations. By introducing a binary association index $x_u = \{0,1\}$ and fixing the number of macro users temporally as $|M_s| = k = \sum_{u=1}^{U} x_u$, where U is the total number of users within the partitioned area, the sub-problem $R_s$ can be formulated into combinatorial optimization problem as follows

$$R_s = \sum_{u \in M_s} \min\left(\frac{\alpha_s W_M C_{u,M}^e}{k}, L_u\right) + \sum_{u \in s} \min\left(\frac{W_s C_{u,s}^e}{U-k}, L_u\right)$$
$$= \sum_{u=1}^{U} \min\left(\frac{\alpha_s W_M C_{u,M}^e}{k}, L_u\right) x_u + \sum_{u=1}^{U} \min\left(\frac{W_s C_{u,s}^e}{U-k}, L_u\right)(1-x_u)$$
$$= \sum_{u=1}^{U} \left(\min\left(\frac{\alpha_s W_M C_{u,M}^e}{k}, L_u\right) - \min\left(\frac{W_s C_{u,s}^e}{U-k}, L_u\right)\right) x_u$$
$$+ \sum_{u=1}^{U} \min\left(\frac{W_s C_{u,s}^e}{U-k}, L_u\right)$$
$$= \mathbf{f}_k^T \mathbf{x}_k + A_k$$
(19)

where $\mathbf{f}_k$, $\mathbf{x}_k$ are vectors whose elements are the objective function value and association index respectively, $A_k$ is a constant value, and $(\cdot)^T$ denotes the transpose operation. Since the number of macro users is fixed, there is an implicit constraint

$$\mathbf{1}^T \mathbf{x}_k = k,$$ (20)

where $\mathbf{1}$ is the vector that all elements are one. Therefore the optimization problem is formulated as follows

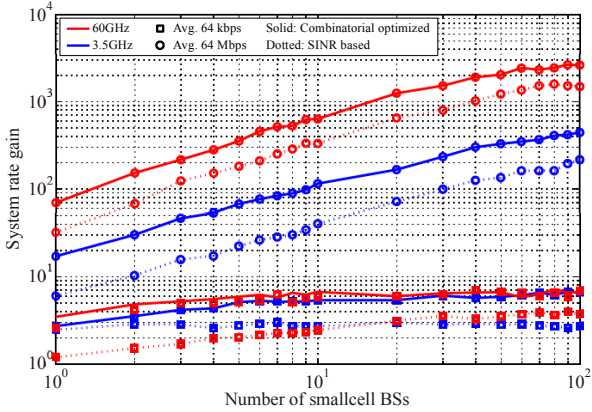
**Fig. 6** Simulation result.

$$\text{maximize } \mathbf{f}_k^T \mathbf{x}_k$$
$$\text{subject to } \mathbf{1}^T \mathbf{x}_k = k \quad (21)$$

This solution is for $k$ user case therefore the optimum solution should be found among $k = 0$ to U.

$$\mathbf{x}^* = \arg\max_{\mathbf{x}_k, k \in \{0,1,\ldots U\}} \mathbf{f}_k^T \mathbf{x}_k \quad (22)$$

### 4.4 Performance Evaluation

**Table 3** Simulation parameters.

| Parameter | Value |
|---|---|
| Bandwidth (Macro / 3.5 GHz / 60 GHz) | 10 MHz / 100 MHz / 2 GHz |
| Number of macro cells | 7 (1 evaluate, 6 interference) |
| Number of small cells (per 1 macro cell) | 0 – 100 |
| Number of UEs (per 1 macro cell) | 5000 |
| Number of BS antennas (Macro / 3.5 GHz / 60 GHz) | 4 / 4 / 1 |
| Number of UE antennas | 2 |
| Macro ISD | 500 m |
| Antenna beam pattern (Macro / 3.5 GHz / 60 GHz) | 3GPP sector [41]/ Omni / WiGig [59] |
| Antenna gain (Macro / 3.5 GHz / 60 GHz) | 17 dBi / 5 dBi / 25 dBi |
| BS antenna height (Macro / 3.5 GHz / 60 GHz) | 25 m / 10 m / 3 m |
| UE antenna height | 1.5 m |
| Tx power (Macro / 3.5 GHz / 60 GHz) | 46 dBm / 30 dBm / 10 dBm |
| Pathloss model [41] [32] (Macro / 3.5 GHz / 60 GHz) | $128.1 + 37.6 \log_{10}(d[\text{km}])[\text{dB}]$ / $140.7 + 36.7 \log_{10}(d[\text{km}])[\text{dB}]$ / Eq. (13) |
| Channel model (Macro / 3.5 GHz / 60 GHz) | 3GPP SCME urban macro scenario [58]/ 3GPP SCME urban micro scenario [58]/ Measurement based Rician fading model [32] |
| Noise power density | -174 dBm/Hz |
| Average traffic demand | 64 kbps / 64 Mbps |

By using the traffic model in Sect. 4.1, the measurement based mm-wave channel model in Sect. 4.2, and the cell association scheme in Sect. 4.3, and, the performance of multi-band HetNet with mm-wave access is evaluated via system level simulation. The evaluation metric is system rate gain which is defined as the gain of system rate compared with the homogeneous network, so that the goal of the analysis is to achieve the system rate gain of 1000 by introducing smallcell BSs. The system rate gain is evaluated by changing the number of smallcell BSs of 3.5 GHz or 60 GHz per macro cell. As for the propagation model, standard 3GPP model [41] is employed for macro and 3.5 GHz small cell links, while measurement based model described in the previous section is employed for mm-wave small cell link. It is assumed that OFDM based transmission scheme is applied in all bands and perfectly works, so that frequency selective fading can be converted into frequency flat fading in each frequency bin. Moreover, as very narrow beam antenna is employed in mm-wave link, the corresponding delay spread is very small in the mm-wave channel. Regarding the MIMO antenna configuration, current standard of 4x2 single user MIMO transmission is considered both for macro and 3.5 GHz small cell links, while 1x2 SIMO transmission is employed for mm-wave as the baseline since rich scattering environment cannot be expected for mm-wave channel. The performance is evaluated by changing the average traffic demands of 64 kbps and 64 Mbps corresponding to present and 10 years later respectively. The remaining simulation parameters are listed in Table 3 where almost all parameters are based on the 3GPP standard for macro and 3.5 GHz small cell links or IEEE 802.11ad standard for mm-wave small cell links.

Numerical results of the system rate gain are shown in Fig. 6 where the horizontal axis shows the number of deployed smallcell BSs and the vertical axis shows the system rate gain. The blue lines show the results with 3.5 GHz small cells while red lines show those of mm-wave. Two different cell association schemes of the combinational optimization and the SINR based algorithm are shown by solid and dotted lines respectively.

First of all, the superiority of the novel cell association scheme with the combinational optimization is clear in all cases, where it can provide twice higher gain in all cases. These results show that proposed cell association method is effective for multi-band HetNets with C-RAN architecture. As a next step, we will compare the performance of 3.5 GHz and 60 GHz. In the present traffic load, there is no significant difference in the system rate gain between two different bands, and about 7 times higher system rate is obtained by introducing HetNets. In 10 years later, as the traffic is very heavy, the performance of smallcell BSs is much higher. The system rate gain will reach 2000 times in the case of 60 GHz smallcell BSs and 400 times for the case of 3.5 GHz smallcell BSs respectively. These results show the feasibility of introducing mm-wave bands to 5G system to increase the system rate by more than 1000 times. Of course, the comparison results highly depend on the simulation parameters. If more bandwidth is available in 3.5 GHz, the system rate gain will be scaled in proportion to the available bandwidth. However, since the system is already interference limited, additional antenna gain will not change the results so much.

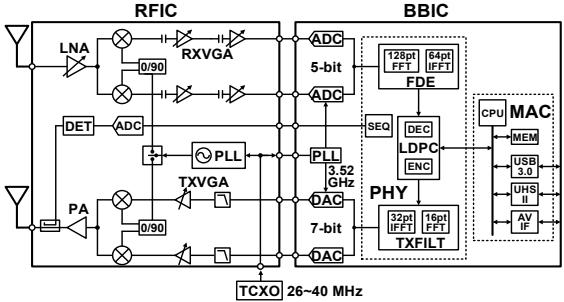

**Fig. 7** Block diagram of RFIC.

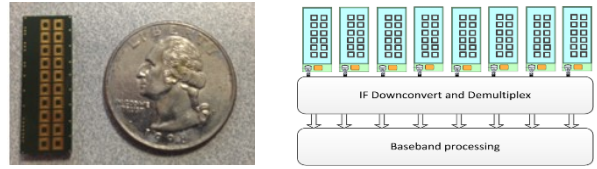

**Fig. 10** Single MAA element (left) and schematics of an 8-module MAA architecture.

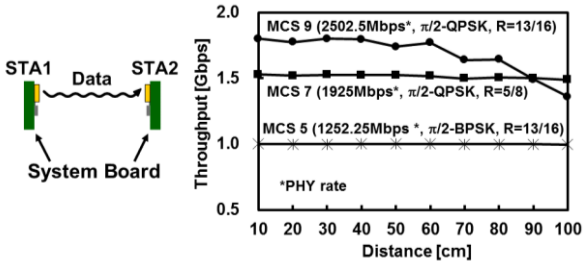

**Fig. 8** RF antenna module and system board.

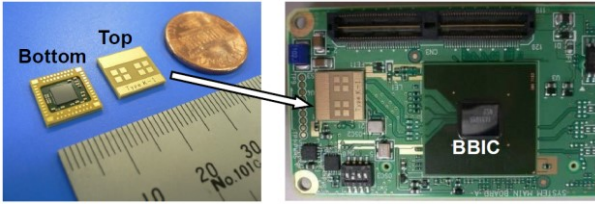

**Fig. 9** Measured MAC throughput over the air.

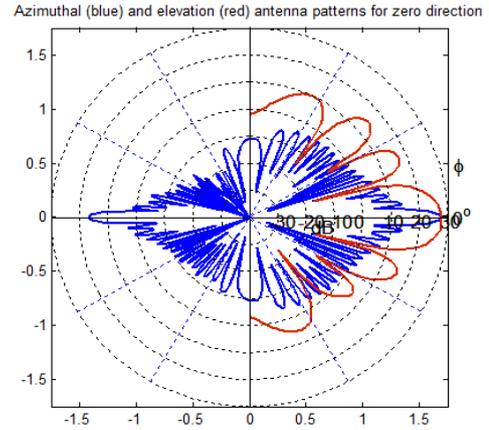

**Fig. 11** 2D antenna patterns for 8-module MAA.

## 5. Millimeter-wave Devices & Regulations

Section 4 showed the feasibility of the proposed 5G cellular networks with mm-wave access via system level simulation by introducing the expected future traffic demands, the new mm-wave channel model, and the novel centralized cell association algorithm. This section on the other hand studies about the feasibility of mm-wave access by giving the status of latest devices for mm-wave communications and mm-wave regulations.

### 5.1 Millimeter-wave Device for Low Power Consumption

This sub-section presents a state-of-the-art low power CMOS transceiver based on the WiGig/IEEE 802.11ad standard. Even though recent works have realized 60 GHz transceivers in a cost-effective CMOS process [42, 43, 44], achieving low power consumption as well as small form factor remains a difficult challenge. By employing sophisticated built-in self-calibrations, the developed chipset achieves MAC throughput of 1.8 Gbps while dissipating less than 1 W total power.

Figure 7 shows the block diagram of the transceiver [45]. The RFIC employs direct conversion architecture, supporting all four channels allocated at 60 GHz. The BBIC includes PHY (PHYsical) and MAC (Media Access Control) layers as well as high speed interfaces. The chipset is developed for single-carrier (SC) modulation, which is suitable for reduced power consumption as compared to OFDM modulation. To overcome performance degradations due to in-band amplitude variations, which are primarily a result of gain variations of analog circuits and multipath delay spread, the chipset employs built-in Tx in-band calibration and an Rx frequency domain equalizer (FDE) [45]. These techniques relax the requirement of the gain flatness and process variations for high speed analog circuits, leading to less power consumption with minimum hardware overhead.

Figure 8 shows the photograph of an RF module and a system board. The RF module employs a cavity structure with the RFIC mounted by flip chip technology. Each Tx/Rx antenna consists of four patch elements, providing 6.5 dBi gain with 50 deg beam width. The RFIC and the BBIC are fabricated in 90 nm CMOS and 40 nm CMOS respectively.

In the Tx mode, the chipset consumes 347 mW in the RFIC and 441 mW in the BBIC with the output power of +8.5 dBm EIRP. In the Rx mode, it consumes 274 mW in the RFIC and 710 mW in the BBIC with 7.1 dB noise figure. Figures 9 shows the measured MAC throughput from one station to the other using different modulation and coding schemes (MCS). The chipset achieves 1.8 Gbps up to 40 cm by using MCS 9 (modulation scheme: π/2-QPSK, coding rate: 13/16) and 1.5 Gbps up to 1 m by using MCS 7 (modulation scheme: π/2-QPSK, coding rate: 5/8).

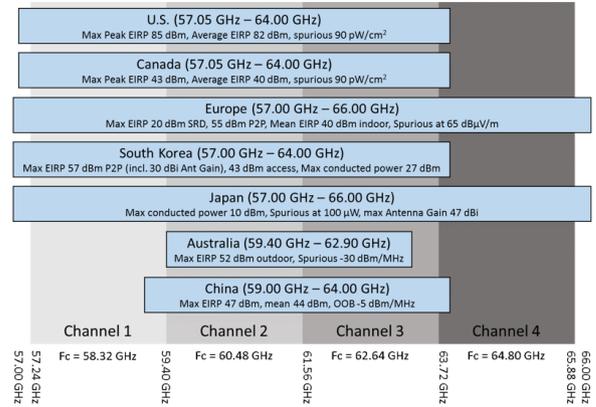

**Fig. 12** Summary of 60 GHz regulations around the globe.

For small cell or backhaul usage, longer communication distance will be required. This is achieved by either increasing the output power or antenna gain. For instance, link margin can be increased by using $N_{Tx}$ or $N_{Rx}$ elements in a phased-array configuration, which can be installed in BSs where size and power constraints are less critical. Ignoring second order effects such as feeding loss from the RFIC to antenna elements, the link budget is increased by

$$10 \log 10(N_{Tx}^2 \cdot N_{Rx}) \quad (23)$$

due to the phased-array gain and the transmitted power increase. As a numerical example, $N_{Tx}$=32 and $N_{Rx}$=4 give 36.1 dB, which translate to 65 times improvement in the communication distance.

5.2 Millimeter-wave Device for High Gain Beamforming

Latest advances in the millimeter wave antenna and packaging technology [46] allow creating the phased antenna arrays but with limited number of elements, due to large losses in the feeding lines. Next evolution in mm-wave technology is modular antenna arrays (MAA) [47, 48], comprised of large number of sub-array modules. Each module has built-in sub-array phase control and coarse beam steering capability. MAA's flexible and scalable architecture accomplishes a wide range of antenna gain and apertures challenging today's regulatory EIRP limits. For example, Fig. 10 left shows one module which may be used for constructing the MAA by any configuration or, as a single phased antenna array, for an UE. The 8-module MAA architecture (each sub-array module is an 8x2=16 elements, vertical x horizontal) and its 2D antenna pattern are shown in Fig. 10 right and Fig. 11 respectively.

Capable of realizing massive MIMO in baseband with independently phase-controlled antenna elements (totally 8x32=128) such MAA is going to increase range up to 400 m for LOS backhaul and mm-wave-capable small cell (MCSC) access range up to 100 m.

First downlink access link (BS 8-module MAA with 19 dBm Tx power, 24 dBi antenna gain, single carrier, π/2-16 QAM modulation, ½ coding rate, and UE with Rx quasi-omni antenna with 5 dBi gain) budget estimates show that a small cell edge throughput of about 3 Gbps for ISD (inter-site distance) 100 m. First uplink access link (BS 32-module MAA with 30 dBi antenna gain and UE with 10 dBm Tx power, quasi-omni antenna with 5 dBi gain, single carrier, π/2-64 QAM modulation, ½ coding rate) budget estimates show that a small cell edge throughput of about 3 Gbps for ISD 100 m. First backhaul link (BS 8-module MAA with 19 dBm Tx power, 24 dBi antenna gain, single carrier, π/2-64 QAM modulation, ½ coding rate at both sides) budget estimates show a highest data rate of 6.5 Gbps at 150 m range.

5.3 Millimeter-wave Regulations

In ITU-R Radio Regulations table of frequency allocations, the so-called 60 GHz band is allocated to a variety of services including Fixed, Mobile, Space Research, and Earth Exploration Satellite (passive) on co-primary basis. The co-primary Mobile allocation spans the entire 57 to 66 GHz range. While not every allocated service is in use around the world, several countries have already included in their regulations provisions for unlicensed use of all or part of 57 to 66 GHz frequency range for multi-gigabit wireless access systems, primarily those adhering to IEEE 802.11ad, or WiGig, standard. Figure 12 summarizes some of the existing regulations in key markets around the globe for 60 GHz band, followed by more detailed information about the United States and European CEPT (Conference of Postal and Telecommunications administrations). References [49, 50, 51, 52, 53, 54, 55, 56, 57] contain detailed information on regulations some of which are reflected in Fig.

12.

In United States, the 60 GHz band is allocated on a co-primary basis to the Federal Mobile, Fixed, Inter-Satellite and Radiolocation services and to non-Federal Fixed, Mobile and Radiolocation services. Recently, the FCC (Federal Communications Commission) modified its rules to allow operation at higher power levels by 60 GHz unlicensed devices that operate outdoors. Specifically, the FCC increased the average/peak EIRP limit to 82/85 dBm minus 2 dB for every dB that the antenna gain is below 51 dBi. With the new rules, it is possible to use large antenna arrays and extend the reach of 60 GHz signals to levels that would sustain a reasonable link budget in distances appropriate for various types of applications including backhaul.

In CEPT, there are several European-wide documents that govern usage of the 60 GHz band for indoor and outdoor applications, including:
- ETSI EN 302 567: "60 GHz Multiple-Gigabit WAS/RLAN Systems; Harmonized EN covering the essential requirements of article 3.2 of the R&TTE directive"
- ERC Recommendation 70-03, "Relating to use of Short Range Devices"
- ECC Recommendation (09)01: "Use of the 57 - 64 GHz frequency band for point-to-point fixed wireless systems"

By summarizing above discussions, the 60 GHz band has potential to be used both for access with beamforming antenna and backhaul world widely. However, to use such a kind of high frequency band for cellular networks with harmonized manner, it is better to standardize such bands, e.g. above 6 GHz, as IMT (international mobile telecommunication) band in ITU-R.

## 6. Conclusion

This paper studied mm-wave evolution for 5G cellular networks to solve the problem of explosion in mobile traffic by introducing mm-wave small cells over the current cellular systems. The overall architecture of 5G cellular networks with mm-wave access was provided, where newly introduced mm-wave smallcell BSs and a conventional macro BS are connected to C-RAN with enhanced CPRI to effectively operate the system via U/C splitting. Different from the previous works, comprehensive system level simulations of the proposed network were conducted by introducing a new expected future traffic model, a measurement based mm-wave propagation model, and a novel centralized cell association algorithm. The realistic simulation results proved the effectiveness of mm-wave access to improve system rate 1000 times higher than the conventional cellular networks in 10 years. It is also found that the performance with mm-wave 60 GHz band is higher than the commonly considered 3.5 GHz band that motivates the engineers to integrate the mm-wave access into future cellular networks. This paper also provided latest mm-wave devices and regulations to prove the feasibility of mm-wave band for the 5G cellular networks.

However, there are several remaining problems of mm-wave to be solved for 5G cellular networks. First of all, the standardization in ITU-R to include higher frequency such as above 6 GHz as IMT bands is needed to spread the mm-wave band in cellular networks world widely. The second is waveform in mm-wave band. By extending the current situation, there are two options. The one is develop interface between 3GPP and Wi-Fi to utilize the current standard of WiGig, and the other is to properly modify LTE waveform to accommodate the mm-wave band in the future developed new RATs. We need to study about the pros and cons of these options and to make consensus. The third problem is CAPEX/OPEX (capital expediter/operating expediter) to introduce such high density small cells. We may need extra new concept such as virtual operator to reduce the cost by sharing the smallcell BSs to make the 5G cellular networks sustainable.

## Acknowledgments

This research has been done as a project named "Millimeter-Wave Evolution for Backhaul and Access (MiWEBA)" under international cooperation program of ICT-2013 EU-Japan supported by FP7 in EU and MIC in Japan, and also this work was partly supported by "The research and development project for expansion of radio spectrum resources" of The Ministry of Internal Affairs and Communications in Japan.

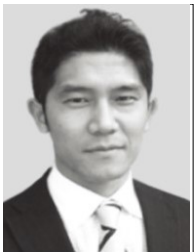

**Kei Sakaguchi** received the B.E. degree in electrical and computer engineering from Nagoya Institute of Technology, Japan in 1996, and the M.E. degree in information processing from Tokyo Institute of Technology, Japan in 1998, and the Ph.D. degree in electrical and electronic engineering from Tokyo Institute of Technology in 2006. From 2000 to 2007, he was an Assistant Professor at Tokyo Institute of Technology. Since 2007, he has been an Associate Professor at the same university. Since 2012, he has also joined in Osaka University as an Associate Professor, namely he has two positions in Tokyo Institute of Technology and Osaka University. He received the Young Engineering Awards from IEICE and IEEE AP-S Japan Chapter in 2001 and 2002 respectively, the Outstanding Paper Award from SDR Forum and IEICE in 2004 and 2005, respectively, the Tutorial Paper Award from IEICE Communication Society in 2006, and the Best Paper Awards from IEICE Communication Society in 2012 and 2013. His current research interests are 5G cellular networks, sensor networks, and wireless energy transmission. He is a member of IEEE.

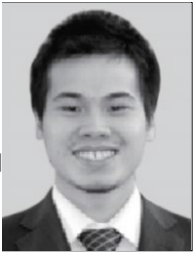
**Gia Khanh Tran** was born in Hanoi, Vietnam, on February 18, 1982. He received the B.E. , M.E. and D.E. degrees in electrical and electronic engineering from Tokyo Institute of Technology, Japan, in 2006, 2008 and 2010 respectively. Currently, he is working as an Assistant Professor at the same university. He received IEEE VTS Japan 2006 Young Researcher's Encouragement Award from IEEE VTS Japan Chapter in 2006 and the Best Paper Award in Software Radio from IEICE SR technical committee in 2009 and 2013. His research interests are MIMO transmission algorithms, multiuser MIMO, MIMO mesh network, wireless power transmission, coordinated cellular networks and mm-wave communications. He is a member of IEEE.

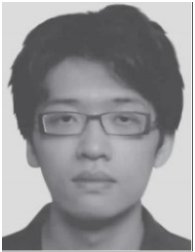
**Hidekazu Shimodaira** was born in Tokyo, Japan, on September 22, 1988. He received the B.E. and M.E. degree in electrical and electronic engineering from Tokyo Institute of Technology, Japan, in 2012 and 2014 respectively. From 2014, he is a Ph. D course student in the Department of Electrical and Electronic Engineering, Tokyo Institute of Technology. His current research interest is heterogeneous and cooperative cellular networks and mm-wave communications. He is a student member of IEICE.

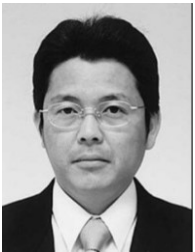
**Shinobu Nanba** received the B.E., M.E., and Ph.D. degrees, in information science and electrical engineering from Kyushu University, Fukuoka, Japan, in 1994, 1996, and 2006, respectively. He joined Kokusai Denshin Denwa Co., Ltd. (now KDDI Corp.) in 1996. His current research interests include cell planning and propagation for cellular mobile systems. He is currently a research engineer at Wireless Platform Laboratory of KDDI R&D Laboratories, Inc. He received the Young Researcher's Award from the Institute of Electronics Information, and Communication Engineers (IEICE) in 2002. He is a member of the IEICE.

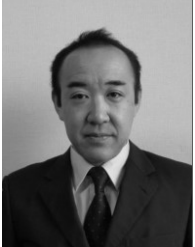
**Toshiaki Sakurai** received the B.S. and M.S. degrees in Electrical Engineering from the Tohoku University in 1996 and 1998. He now with AVC Networks company, Panasonic Corporation. He has mainly been involved research and product development work of mobile communication protocol, wireless access methodology, resource management.

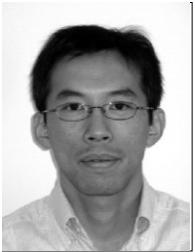
**Koji Takinami** received the B.S. and M.S. degrees in electrical engineering from Kyoto University, Kyoto, Japan, in 1995 and 1997, respectively, and the Ph.D. degree in physical electronics from Tokyo Institute of Technology, Tokyo, Japan, in 2013. In 1997, he joined Matsushita Electric Industrial (Panasonic) Co., Ltd., Osaka, Japan. Since then he has been engaged in the design of analog and RF circuits for wireless communications. From 2004 to 2006, he was a visiting scholar at the University of California, Los Angeles (UCLA), where he was involved in the architecture and circuit design of the high efficiency CMOS power amplifier. In 2006, he joined Panasonic Silicon Valley Lab, Cupertino, CA, USA, where he worked on high efficiency transmitters and low phase-noise digital PLLs. In 2010, he relocated to Japan and currently leads the development of the millimeter wave transceiver ICs. Dr. Takinami is a co-recipient of the Best Paper Award at the 2012 Asia-Pacific Microwave Conference. He has been a member of the IEEE ISSCC Technical Program Committee since 2012.

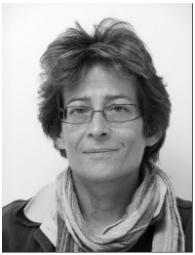
**Isabelle Siaud** received the Electronic Master Dipl from the University Pierre et Marie Curie (UPMC Paris VI) in 1992 and received a multi-disciplinary (mm-wave propagation and short range PHY/MAC system design) PhD from the INSA-Rennes in 2011 prepared as R&D engineer in Orange. From 1993 to 1998, she participated to propagation modelling for 3G and short range mm-waves. In 1999, she joined the Orange labs in Renne in charge of defining innovative PHY/MAC layer systems covering mm-wave UWB- OFDM extensions, Multiple Interface Management and green transmission techniques for future 4G and 5G networks. She has been a head of the UWB-MultiCarrier (UWB-MC) cluster in the IST/FP6 MAGNET project and contributed to several collaborative projects (IST-FP7 IPHOBAC, ICT-FP7 OMEGA, MAGNET Beyond, etc...) dealing with innovative 60 GHz transmissions (radio, RoF) and interface management. Until 2010, she co-supervised the "Short range WG" of theWWRF. Actually, she is involved in the GreenTouch consortium, the ICT-FP7 METIS and the EU/Japan MiWEBA projects. MiWEBA evaluates mm-overlay heterogeneous networks for 5G. She recently joined the BCOM Technical Research Institute in Rennes, in charge of elaborating new research collaborative projects related to future of hypermedia and smarter networks.    She is co-author of approximately 40 international conferences, 3 books and 8 patents.

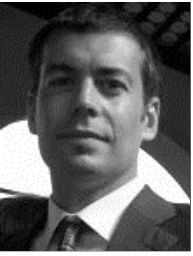

**Emilio Calvanese Strinati**   obtained his Engineering Masters degree in 2001 from the University of Rome 'La Sapienza' and his Ph.D in Engineering Science in 2005 on Radio link control for improving the QoS of wireless packet transmission. He then started working at Motorola Labs in Paris in 2002. Then in 2006 he joint CEA/LETI as a research engineer. From 2007, he becomes a PhD supervisor. Since 2011 he is the Smart Devices & Telecommunications European collaborative strategic programs Director. E. Calvanese Strinati has published around 70 papers in international conferences and books chapters, and is the main inventor or co-inventor of more than 50 patents. He has organized more than 20 international workshops and special sessions on green communications and heterogeneous networks hosted in international conferences as IEEE GLOBCOM, IEEE PIMRC, IEEE WCNC, IFIP, and European Wireless. Dr. Calvanese Strinati he is the project manager in CEA of the Green Communication EARTH IP-FP7 project and he has been the co-chair of the wireless working group in GreenTouch from April 2010 to January 2012. Since 2012 he is the strategy director of the Smart Devices & Telecommunications Strategy Program Director and in 2013 he has been elected as one of the 5G PPP steering board members.

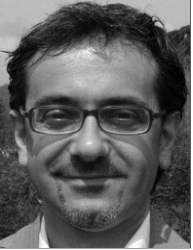

**Antonio Capone**   is Full Professor at Politecnico di Milano (Technical University), where he is the director of the Advanced Network Technologies Laboratory (ANTLab). His expertise is on networking and his main research activities include protocol design (MAC and routing) and performance evaluation of wireless access and multi-hop networks, traffic management and quality of service issues in IP networks, and network planning and optimization. On these topics he has published more than 200 peer-reviewed papers in international journal and conference proceedings. He received the M.S. and Ph.D. degrees in electrical engineering from the Politecnico di Milano in 1994 and 1998, respectively. In 2000 he was visiting professor at UCLA, Computer Science department. He currently serves as editor of ACM/IEEE Trans. on Networking, Wireless Communications and Mobile Computing (Wiley), Computer Networks (Elsevier), and Computer Communications (Elsevier). He serves often in the TPC of the major conferences of the networking research community and in their organizing committees. He is a Senior Member of the IEEE.

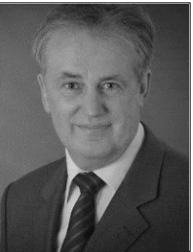

**Ingolf Karls**   received his B.S. and M.S. degrees in Electrical Engineering from Technical University Chemnitz 1985 and 1997, respectively. He contributed to several generations of wireless communication products at Siemens AG and Infineon Technologies AG. He is currently at Intel Mobile Communications working at millimeter wave systems and 5G.

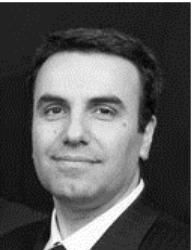

**Reza Arefi**   leads spectrum strategies and global radiocommunication standards at Intel's Standards & Advanced Technology group at Intel Mobile Communications Group. He develops market-driven spectrum strategies for Intel's wireless products and assists with global public policy efforts. With over twenty years in telecom industry, Reza has been actively contributing to standards and various industry groups on wireless systems for the past fourteen years, often in leadership positions (ITU-R, IEEE, WiGig Alliance, etc.). Since 2004, Reza has regularly represented Intel in regional and international regulatory standards organizations. He has also acted as Intel delegate to 2012 World Radiocommunication Conference (WRC-12). His cur-rent focus is on enabling spectrum for 5G cellular systems. He holds several patents in various areas of wireless communications including mm-wave.

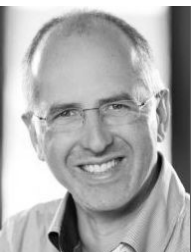

**Thomas Haustein**   received his the Dr.-Ing. (Ph.D.) degree in mobile communications from the Technische Universitat Berlin, Germany, in 2006. In 1997, he joined the Fraunhofer Heinrich Hertz Institute, Berlin, where he worked on wireless infrared systems and radio communications with multiple antennas and orthogonal frequency division multiplexing. He focused on real-time algorithms for baseband processing and advanced multiuser resource allocation. Since 2006, he was with Nokia Siemens Networks, where he conducted research for Long-Term Evolution (LTE) and LTE-Advanced. Since 2008, he is the Head of the Wireless Communications and Networks Department at Fraunhofer HHI. His research expertise is in the area of MIMO, OFDM, LTE-Advanced, Cognitive Radio Systems and Radio technology candidates for future wireless systems - 5G. Thomas is academic advisor to NGMN and actively contributing to industry consortia like IWPC and Green Touch.